\documentclass[prl,twocolumn,showpacs]{revtex4}
\usepackage{amsmath}
\usepackage{amsfonts}
\usepackage{amssymb}
\usepackage{graphicx}
\usepackage{color}
\usepackage{textgreek}

\begin{document}

\title{Chip-based squeezing at a telecom wavelength}

\author{F. Mondain$^1$ ,T. Lunghi$^1$, A. Zavatta$^{2,3}$, E. Gouzien$^1$, F. Doutre$^1$, M. De Micheli$^1$, S. Tanzilli$^1$, V. D'Auria$^1$}

\email{virginia.dauria@univ-cotedazur.fr}
\affiliation{$^1$Universit\'e Cote d'Azur, CNRS, Institut de Physique de Nice,  Parc Valrose, 06108 Nice Cedex 2, France\\
$^2$Istituto Nazionale di Ottica (INO-CNR) Largo Enrico Fermi 6, 50125 Firenze, Italy\\
$^3$LENS and Department of Physics, Universit\'a di Firenze, 50019 Sesto Fiorentino, Firenze, Italy}




\begin{abstract}

We demonstrate a squeezing experiment exploiting the association of integrated optics and telecom technology as key features for compact, stable, and practical continuous variable quantum optics. In our setup, squeezed light is generated by single pass spontaneous parametric down conversion on a lithium niobate photonic circuit and detected by an homodyne detector whose interferometric part is directly integrated on the same platform. The remaining parts of the experiment are implemented using commercial plug-and-play devices based on guided-wave technologies. We measure, for a CW pump power of 40\,mW, a squeezing level of $-2.00\pm0.05$\,dB,  (antisqueezing  $+2.80\pm0.05$\,dB) thus confirming the validity of our approach and opening the way toward miniaturized and easy-to-handle continuous variable based quantum systems. 
\end{abstract}


\maketitle

\section{Introduction}

Over the last years, single and  multimode squeezed states~\cite{Lvovsky2014} have played a crucial role in the development of quantum technologies such as quantum computation~\cite{Leuchs2010, Pfister2007}, communication~\cite{Werner2013,  Pirandola2012, Peng2018}, and sensing~\cite{Schnabel2017}. In this context, important experimental and theoretical tools for squeezing generation, manipulation, and analysis have been developed~\cite{Leuchs2016, Lvovsky2009}. Moreover, it has been proved that squeezed states can be manipulated so as to generate highly non-classical states, such as Schr{\"o}dinger kittens~\cite{Alexei2006} or cats~\cite{Lvovsky2017} or hybrid entangled states featuring continuous-discrete variables properties~\cite{Julien2014, LvovskyHib2017}. \\
So far, most of important squeezing demonstrations have been performed by exploiting bulk optics experiments, where squeezed states are usually generated via resonant systems, such as optical parametric oscillators (OPO)~\cite{Schnabel2011, Treps2013}, and detected thanks to free-space homodyne detections, for which careful spatial alignment and mode matching are required~\cite{Furusawa2015}. As a consequence, in view of practical applications, we have been assisting to the miniaturisation of important building blocks of squeezing experiments. Compact and stable squeezing generation has been reported in single-pass waveguides~\cite{Furusawa2007, Hirano2008} and, more recently, in OPO-like devices such as a silicon micro-ring~\cite{Dutt2015} or  a waveguide cavity resonator on lithium niobate allowing to measure a squeezing of $-2.9$\,dB for a pump power of $\approx $23\,mW~\cite{Silberhorn2017}. In parallel, other realisations have reported on-chip homodyne detectors exploiting integrated optics on different substrates~\cite{Furusawa2015, Matteo2018, Raffaelli2018}. Eventually, in the last months, a photonic platform for continuous variable quantum information has been demonstrated on lithium niobate~\cite{Lobino2018}.\\
\begin{figure*}[htbp]
\centering
\includegraphics*[width=1\columnwidth]{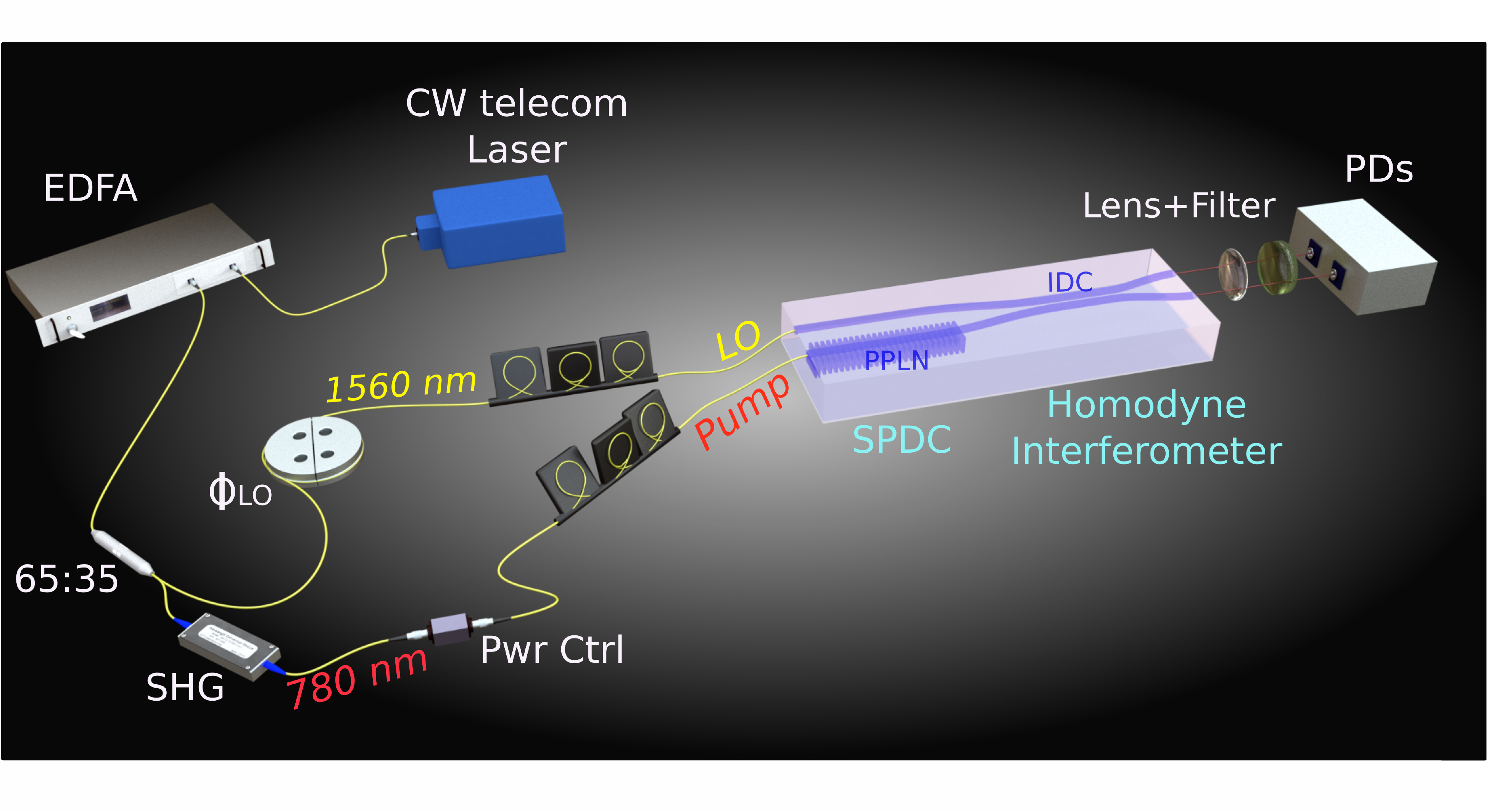}
\caption{Experimental setup. A fibre coupled CW telecom laser at 1560.44\,nm is amplified (EDFA) and split into two by means of a 65:35 fibre beam splitter (f-BS). The less intense output (upper arm) serves as local oscillator (LO) while the brighter one (lower arm) is frequency doubled via SHG in a PPLN/RW and used to pump a on-chip squeezing generation stage (SPDC). The power of the beam at 780.22\,nm is controlled with an in-line fibred attenuator (Pwr Ctrl) and its polarisation is adjusted by means of a fibre polarisation controller. At the output of the SPDC stage, squeezed light at 1560.44\,nm is optically mixed with the LO beam inside the same chip in an integrated directional coupler realizing the interferometric part of the homodyne detector. At the chip output, after passing through a bulk lens followed by an optical filter suppressing residual pump at 780.22\,nm, light is sent to two InGaAs photodiodes (PDs). The LO phase is scanned thanks to a home-made fibre-stretcher ($\phi_{LO}$).}
\label{setup}
\end{figure*}
Our work follows this emerging and very exciting research line. It addresses the realisation of stable, compact, and telecom-compliant squeezing experiments by merging integrated optics on lithium niobate~\cite{Alibart16} and mature classical technologies. This association allows satisfying in a simple way the requirements of easy-to-use and scalable experiments in view of out-of-the-laboratory quantum communication in optical fibres~\cite{Peng2018, Peng2011}. The squeezing generation and the optical coupler for the homodyne detector are fully integrated on a single chip with no significant loss between these two stages and no spatial mode matching concerns~\cite{Furusawa2015}. At the same time, with the only exception of homodyne photodiodes, all the other building blocks of our experiment are realised by means of off-the-shelf guided-wave telecom components, fully compatible with existing fibre networks and allowing fast and plug-an-play experiment reconfigurations~\cite{Optica2016}. More in details, squeezed light is generated on-chip via efficient single-pass spontaneous parametric down conversion (SPDC) in a periodically poled waveguide. Squeezing is emitted in the telecom C-band of wavelengths so as to be compatible with low propagation losses in optical fibres. Its homodyne detection, exploiting on-chip optical mixing with the local oscillator (LO) beam, exhibits low losses, essentially due to Fresnel reflection at the chip end-facet and  photodiode quantum efficiency. Thanks to this approach, we measure with the integrated homodyne interferometer a raw single-mode squeezing level of $-2.00\pm0.05$\,dB ($-3.00\pm0.05$\,dB corrected by avoidable losses), with a continuous wave (CW) pump power of 40\,mW. This value validates our approach and represents, to our knowledge, the best squeezing level obtained in miniaturized systems in single-pass CW pumping regime ~\cite{Lobino2018, Furusawa2007, Silberhorn2017}.

\section{Experimental setup and photonic chip}

\begin{figure}[htbp]
\centering
\fbox{\includegraphics[width=0.7\columnwidth]{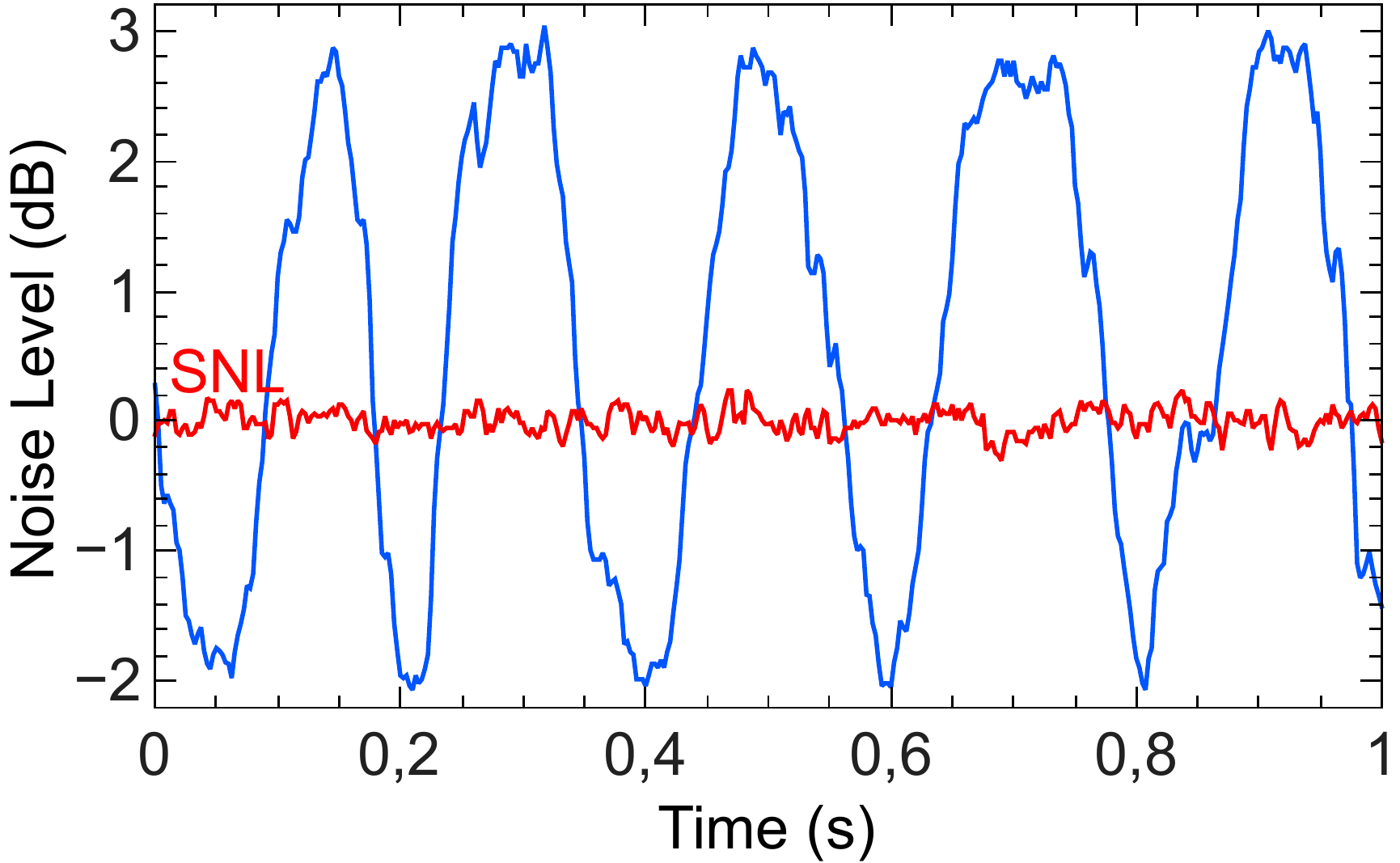}}
\caption{Schematic of our photonic chip on lithium niobate. The chip includes an SPDC stage, consisting in a periodically poled waveguide (3\,cm long with poling period $\Lambda=16.3~\mu$m) for squeezing generation at 1560.44\,nm, and an integrated directional coupler realizing the interferometric part of the homodyne squeezing detection. The whole chip length is 5\,cm. All waveguide are obtained by soft proton exchange\,\cite{Chanvillard2000} and have a width of 6~\textmu m. The 127~\textmu m spacing between the input (output) waveguides is compatible with of-the-shelf fibre-arrays. The homodyne photodiodes are outside the chip and are bulk commercial components.}
\label{chippo}
\end{figure}

The experimental setup is sketched in Fig.\,\ref{setup}. It associates an injection system, exploiting easy-to-assemble guided-wave components, with our home-made lithium-niobate chip. \\
The setup relies on a master, fibre-coupled laser generating a CW optical beam at 1560.44\,nm and amplified up to 0.95\,W with an erbium-doped fibre amplifier (EDFA). At the output of the EDFA, the single mode laser light is split by a high-power 65:35 fibre coupler (f-BS). The less intense output is directed to a home-made fibre stretcher allowing to scan its phase, to be subsequently used as local oscillator for the homodyne interferometer. 
The brighter output is frequency doubled to 780.22\,nm via second-harmonic generation (SHG) and used to pump the squeezer. The single pass SHG is realised in a commercial periodically poled lithium-niobate ridge waveguide (PPLN/RW~\cite{NTTridge}) where both input and output ports are fibre-coupled. We note that the ridge output coupling is optimised by the manufacturer so as to maximise the collection only at 780.22\,nm. Residual light at 1560.44\,nm at the ridge output is suppressed during its propagation in visible light single mode fibres and by a fibre wavelength demultiplexer (WDM 980/1550, not represented in Fig.\,\ref{setup}).\\ 
The LO at 1560.44\,nm and the pump at 780.22\,nm are sent to a fibre array and  butt-coupled to the home-made photonic chip. A fibre polarisation controller (PC) on each arm is introduced to properly adjust the polarisation. Typical fiber-to-input guide couplings are of $\approx 0.60$. This value can be increased up to $0.92$ by inserting taper structures on the photonic chip input \cite{Castaldini2007}. The entire optical setup upstream the chip is made using commercial plug-and-play components, exploiting telecom and nonlinear optics technologies and guaranteeing a quick operation with no spatial alignment procedure~\cite{Furusawa2015, Optica2016}.\\
 A schematic of the photonic circuit is presented in Fig.\,\ref{chippo}. The 5\,cm-long chip integrates on a single congruent lithium niobate substrate the two key components for  squeezing experiments, namely the squeezing generation stage and the optical coupler required for the interferometric part of the homodyne detection. Accordingly, it has two input ports, one for the squeezer pump and the other for the homodyne LO, and two output ports that are directly connected to the bulk homodyne photodiodes. The separation between ports on the same facet is 127\,\textmu m, \emph{i.e.} compatible with standard fibre-array coupling systems.  The waveguide structures are 6\,$\mu$m wide and they are fabricated using soft-proton exchange (SPE) following the technique discussed in Ref.~\cite{Chanvillard2000}. Propagation losses have been measured to be $\leq$0.04~dB/cm at 1560\,nm and $\sim$1~dB/cm at 780\,nm. \\
Lithium niobate is particularly suitable to develop integrated squeezers and photonic circuits, featuring high nonlinearity and the possibility of implementing quasi-phase-matching to engineer the nonlinear optical response~\cite{Alibart16}. Our on-chip squeezer is a straight periodically poled waveguide, 3\,cm long and designed to have, when pumped at 780.22\,nm, a type-0 frequency degenerate SPDC centered at 1560.44\,nm. Periodic poling pattern is created by standard electric-field assisted technique with a period $\Lambda$ = 16.3~\textmu m~\cite{Alibart16}. The quasi phase-matching temperature for reaching the desired interaction is 104$^{\circ}$C. Such a high temperature has been chosen so as to mitigate light-induced local modifications of refractive index in lithium niobate waveguides. Photorefractive effects are, indeed, particularly evident for optical signals at shorter wavelengths and, at high powers, can induce hopping between different spatial modes ~\cite{Glass1975}. At the chosen working temperature, no mode-hopping is observed for red powers up to 40\,mW. Moreover, by exploiting on-chip residual cavity effects and evaluating the resonance positions as functions of the input red power, refractive index variations have been measured to be $\leq4\times10^{-5}$. These very small values are consistent with other reported results~\cite{Hellwig11}. The quality of the SPDC stage has been assessed by means of single-photon counting technique. Typical values for our conversion efficiencies are $\sim 10^5$ photon pairs/mW/GHz/s with a FWHM for the spectral emission of 70\,nm~\cite{Lutfi2015}. We recall that, due to the absence of optical resonator, this value corresponds directly to the squeezing bandwidth~\cite{Furusawa2007, Pfister2009}. \\
The on-chip homodyne interferometer is based on a balanced integrated directional coupler (IDC), exploiting evanescent tail coupling~\cite{Barral2015}. It consists of two waveguides running close to each other over a geometrical length of $\sim$5.5\,mm and with a centre-to-centre separation of 11~\textmu m. The two IDC inputs are fed with the squeezer output (directly on-chip) and with the local oscillator. Its measured splitting ratio is 50:50 at the SPDC central emission wavelength (\emph{i.e.} at the LO wavelength). To achieve such a good balancing, the ideal coupling length and BS design have been numerically computed based on our waveguides' typical measured properties. Moreover, in order to comply with fabrication uncertainties, we have realised on the same sample 17 copies of our photonic circuit, corresponding each to a slightly different IDC length, numerically calculated on the basis of our usual parameter fluctuations. The optimal component has been selected among these copies.\\
The IDC outputs are directed outside the chip and sent to two bulk photodiodes as required for the homodyne detection. The chip is diced using a semiconductor saw and polished at 0$^{\circ}$ through chemical mechanical polishing. Due to the absence of anti-reflection coating, light experiences, at its output, a sharp refractive index change at the lithium niobate-air interface and is transmitted with an efficiency $\eta_{F}=1-R_{Fresnel}$, where $R_{Fresnel}$=$(\frac{n_{air}-n_{chip}}{n_{air}+n_{chip}})^2$. At telecom wavelength, $\eta_F\approx0.86$~\cite{Zelmon1997}. Thanks to a bulk C-coated lens with 11\,mm focal length, the two beams coming out of the photonic circuit are directly imaged on two InGaAs photodiodes (PDs) exhibiting a quantum efficiency $\eta_{PD} \approx$ 0.88 at 1560~nm. By doing so, a nearly perfect chip-to-photodiode coupling is obtained. Residual transmitted light at 780.22\,nm is rejected of more than -40\,dB with a bulk optical filter, exhibiting a transmission $\eta_f$=0.99 at 1560~\,nm. We stress that the lens, the filter and the homodyne photodiodes are the only bulk optics components in our setup.
The difference of the photocurrents from the two photodiodes is amplified by a home-made low-noise trans-impedance amplifier with bandwidth of $\sim$5~MHz. Noise power is directly measured with an electronic spectrum analyser set at zero-span centred at 2~MHz. For 0.5~mW LO power on each photodiode, we obtain an electronic signal-to-noise ratio (SNR) of 12.8 dB. Residual electronic noise effect can be taken into account by introducing an additional loss, through the efficiency $\eta_{e}=(SNR-1)/SNR\approx 0.95$~~\cite{Lvovsky2007}.
\begin{figure}[htbp]
\centering
\fbox{\includegraphics[width=0.7\linewidth]{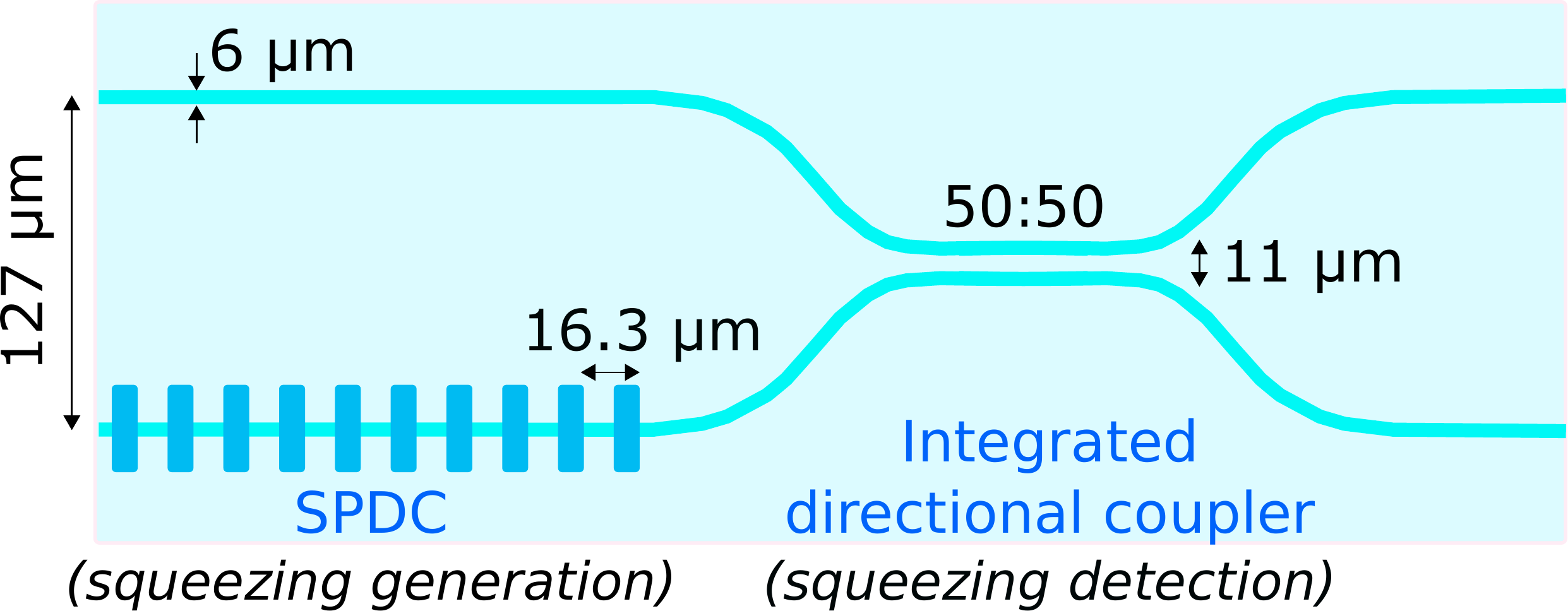}}
\caption{Normalised noise variances at 2\,MHz obtained for a coupled pump power of 40\,mW  as a function of the LO phase (proportional to the time) and with a sweep time of 1\,s. The electrical spectrum analyser resolution and the video bandwidths are 100\,kHz to 30\,Hz, respectively.}
\label{squeezing}
\end{figure}
\section{Experimental results}

Fig.\,\ref{squeezing} shows a typical squeezing curve obtained by scanning the phase of the local oscillator over time. We measure a raw squeezing value of $-2.00\pm0.05$\,dB, with an anti squeezing of $2.80\pm0.05$\,dB when 40\,mW pump light is coupled in the PPLN arm of the chip. The measured squeezed value is affected by losses at the photonic circuit output, such as Fresnel reflection and detectors' non-unit efficiencies as well as electronic noise. However, we underline that these contributions are not due to unavoidable limitations of our chip and can be easily circumvented by means of anti-reflection coating on the chip output facet, leading to $\eta_F\approx$1, and by replacing photodiodes and electronics with more performant ones ($\eta_{PD}\cdot\eta_e\approx$0.99) ~\cite{Schnabel2011}. By correcting our measured squeezing for the overall measurement efficiency, $\eta=\eta_F\cdot\eta_f\cdot\eta_{PD}\cdot\eta_e=0.71$, we can infer the squeezing at the output of the photonic circuit to be $\sim$-3.2\,dB (anti-squeezing of $\sim$3.6\,dB), which is, to our knowledge, the best reported value for CW-pumped squeezing in waveguides without an optical resonator~\cite{Lobino2018, Furusawa2007, Silberhorn2017}. We note that the product of squeezing and anti-squeezing net variances is close to the one expected for  minimum uncertainty states, thus showing the absence of unwanted excess noise on anti-squeezed quadratures. This value provides evidence of the chip quality in terms of both squeezing generation and losses in the interferometric part of the homodyne detector. We underline that our reported squeezing level is compatible with applications in quantum communication protocols like continuous variable entanglement distribution and teleportation \cite{Leuchs2010} as well as with the heralded generation of Schr{\"o}dinger cat-like states \cite{Alexei2006} fully compatible with existing quantum networks and with their use for hybrid entangled states \cite{Julien2014, LvovskyHib2017}. At the same time, we note that the optical chip reported here represents a first important building block for more complex optical circuits able to embrace nonlinear and linear operation stages, enabling new perspectives towards quantum enhanced optical processors \cite{Leuchs2010, Lobino2018}. Eventually, we stress that the squeezing levels can be further improved by increasing the SPDC pump powers. In these regimes, to avoid detrimental photo-refractive in the waveguides, an Mg:O-doped lithium niobate substrate should be used for the photonic chip. Based on our previous results this kind of support \cite{Optica2016}, we expect  for a Mg:O doped SPE waveguide with the same characteristics as the one reported in this paper, a maximum squeezing from $-10$ to $-9$\,dB for an SPDC pump power $\approx500\,mW$. \\

In conclusion, we have demonstrated a compact and easy-to-handle experiment relying on integrated optics on lithium niobate and off-the-shelf telecom and nonlinear components. Our photonic circuit integrates on-chip the squeezing generation and the homodyne interferometer. The squeezed light generated at telecom wavelength exhibits a raw noise compression of $-2.00\pm0.05$\,dB for a pump power of 40\,mW with an anti-squeezing of $+2.80\pm0.05$\,dB. The whole remaining setup employs plug-and-play components requiring no alignment procedures for spatial mode matching. These advantages guarantee an extreme reliability and make our approach a valuable candidate for real-world applications based on continuous variable quantum systems. \\

\textbf{Funding Information}. Agence Nationale de la Recherche ($\emph{Hy-Light}$, ANR-17-CE30-0006-01; $\emph{SPOCQ}$, ANR-14-CE32-0019), Executive Agency for Higher Education, Research, Development and Innovation Funding ($\emph{INQCA}$, PN-II-ID-JRPRO-FR-2014-0013), French government through "Investments for the Future" of Universit{\'e} C\^ote d'Azur UCA-JEDI project (under the label Quantum@UCA, managed by the ANR, ANR-15-IDEX-01). European Regional Development Fund (FEDER) through Project OPTIMAL.\\

\textbf{Acknowledgement}.A. Z. acknowledges CNR Short-Term Mobility Program and the Universit{\'e} Nice Sophia Antipolis for invited professor fellowship. Authors thank H. Tronche for technical support.

\end{document}